\RequirePackage{fix-cm}
\documentclass[smallextended]{svjour3}       
\smartqed  
\usepackage{graphics,epstopdf}
\usepackage{graphicx}
\usepackage[colorlinks=true,citecolor=blue, urlcolor=blue]{hyperref}
\usepackage{float}
\usepackage{amsmath}
\usepackage{latexsym}
\usepackage{amssymb}
\usepackage{graphicx}

\begin{document}

\title{Ontological models, preparation contextuality and nonlocality}



\author{Manik Banik         \and
        Some Sankar Bhattacharya\and
        Sujit K Choudhary\and
        Amit Mukherjee\and
        Arup Roy
}


\institute{Manik Banik \at
              Physics and Applied Mathematics Unit, Indian Statistical Institute, 203 B.T. Road, Kolkata-700108, India\\
              \email{manik11ju@gmail.com}
              \and
           Some Sankar Bhattacharya \at
              Physics and Applied Mathematics Unit, Indian Statistical Institute, 203 B.T. Road, Kolkata-700108, India\\
              \email{somesankar@gmail.com}           
           \and
          Sujit K Choudhary\at
              Institute of Physics, Sachivalaya Marg, Bhubaneswar-751005, Orissa, India\\
                            \email{sujit@iopb.res.in}           
           \and
                     Amit Mukherjee\at
                         Physics and Applied Mathematics Unit, Indian Statistical Institute, 203 B.T. Road, Kolkata-700108, India\\
                                       \email{amitisiphys@gmail.com}
            \and
                      Arup Roy\at
                          Physics and Applied Mathematics Unit, Indian Statistical Institute, 203 B.T. Road, Kolkata-700108, India\\
                                        \email{arup145.roy@gmail.com}                          
           }

\date{Received: date / Accepted: date}

\maketitle

\begin{abstract}
The ontological model framework for an operational theory has generated much interest in recent years. The debate concerning reality of quantum states has been made more precise in this framework. With the introduction of generalized notion of contextuality in this framework, it has been shown that completely mixed state of a qubit is \emph{preparation contextual}. Interestingly, this new idea of preparation contextuality has been used to demonstrate nonlocality of some $\psi$-epistemic models without any use of Bell's inequality. In particular, nonlocality of a non maximally $\psi$-epistemic model has been demonstrated from preparation contextuality of a maximally mixed qubit and Schr\"{o}dinger's steerability of the maximally entangled state of two qubits [\href{http://link.aps.org/doi/10.1103/PhysRevLett.110.120401}{Phys. Rev. Lett {\bf 110}, 120401 (2013)}]. In this paper, we, show that any mixed state is preparation contextual. We, then, show that nonlocality of any bipartite pure entangled state, with Schmidt rank two, follows from preparation contextuality and steerability provided we impose certain condition on the epistemicity of the underlying ontological model. More interestingly, if the pure entangled state is of Schmidt rank greater than two, its nonlocality follows without any further condition on the epistemicity. Thus our result establishes a stronger connection between nonlocality and preparation contextuality by revealing nonlocality of any bipartite pure entangled states without any use of Bell-type inequality.   

\end{abstract}

\section{Introduction}
The nature of quantum state has been debated since the inception of quantum theory (QT) \cite{epr,bohr,popper,ballentine,peres1,caves,pbr}. Does it represent the physical reality or observer's knowledge about the system? In the \emph{ontological models} framework, introduced by Harrigan and Spekkens \cite{harrigan}, this discussion has been made much more precise. An ontological model which aims to reproduce quantum predictions, can be of two types -- $\psi$-ontic or $\psi$-epistemic \cite{harrigan}. The model is said to be $\psi$-epistemic if it considers the quantum mechanical state $\psi$ to represent observer's knowledge about the system. This is in contrast to  $\psi$-ontic view point which considers $\psi$ to represent reality of the system. The de-Broglie-Bohm model \cite{bohm1,bohm2} is an example of this later view point where $\psi$ is given the ontic status. A $\psi$-ontic model can be incomplete, however, in the sense that the quantum mechanical state $\psi$ does not provide the complete description of reality, all by itself. This is the case with the de-Broglie-Bohm model where although $\psi$ is a part of the reality, but it does not provide the complete reality of the system. The complete description is obtained when it is supplemented by the positions of the particles.

Not all the predictions of Quantum theory are compatible with those of 
a \emph{local-realistic} theory. Quantum theory also shows contradiction with a noncontextual
Hidden Variable Theory (HVT). The incompatibility of quantum theory with \emph{local-realism}
was first established by Bell \cite{bell}. On the other hand 
Kochen and Specker (KS) have shown that any realistic interpretation for a quantum system 
described by Hilbert space of dimension greater than two must be \emph{contextual} \cite{kochen}.
The notion of \emph{contextuality} has been generalized recently by Spekkens \cite{spek05} where he 
has demonstrated \emph{preparation contextuality} of a completely mixed state of a qubit.  
In a subsequent work, he, along with Harrigan, has shown that in order to rule out the locality 
of any theory in which $\psi$ has got an ontic status, one does not need Bell's inequality, 
a straightforward argument based on preparation contextuality of completely mixed state of 
a qubit and steerability of maximally entangled state of two qubits suffice \cite{harrigan}. 
The authors of \cite{harrigan} further remark that Bell's inequality is only necessary to 
rule out locality for $\psi$-epistemic HVT. Surprisingly, in a very 
recent development, Liefer and Maroney have demonstrated the nonlocality even 
for a $\psi$-epistemic model without any use of Bell's inequality \cite{maroney} 
provided the model is not \emph{maximally $\psi$-epistemic}.

In this article, we pose the question in a different way: whether this 
approach of revealing nonlocality by preparation contextuality and steerability, 
can be extended to demonstrate nonlocality 
of any bipartite pure entangled state? This question is similar in spirit to the question asked by Nicolas
Gisin in \cite{nicholas}.  Bell 
established the nonlocality of quantum mechanics by means of an inequality which is violated by the singlet state of a pair of qubits. Later, a few other states were discovered which also violate this 
inequality and thus forbid any local-realistic description for them. But, all these 
led to a very natural and very important question; whether this contradiction between QT 
and local-realism is typical or it is restricted to some very special cases. 
In \cite{nicholas}, Gisin answered the question by showing that any pure entangled 
state of a bipartite system violates a Bell's inequality and thus is nonlocal.

We also answer our question affirmatively by showing that the nonlocality of any bipartite pure entangled state, with Schmidt rank greater than two, follows from preparation contextuality of mixed quantum states and steerability
of entangled states \cite{Schro,hjw,ghjw2}. On the other hand, for the entangled states with Schmidt rank two, nonlocality can be demonstrated in the above fashion (i.e. from preparation contextuality steering) provided we impose certain condition on the epistemicity of the underlying ontological model.

The organization of the paper is as follows. In Section-\ref{sec2}, we briefly describe the ontological model framework of an operational theory. Section-\ref{sec3} starts with a brief description of the generalized 
notion of \emph{contextuality}. Here, we adapt the proof of \cite{spek05} to show that any mixed state of a qubit is  preparation contextual \footnote{In \cite{spek05}, preparation contextuality has been shown
for maximally mixed state of a qubit.}. But, this preparation contextuality cannot be directly linked to the nonlocality of entangled states. In this Section, we also discuss the reason behind this.
However, if some extra condition 
is put on the $\psi$-epistemic model, the preparation contextuality can be linked with the nonlocality of any pure bipartite state ---which is the subject matter of Section-\ref{sec4}. 

\section{Ontological model and ontic-epistemic classification}\label{sec2}
We start with briefly discussing the ontological model framework for an operational theory 
as introduced by Harrigan and Spekkens in \cite{harrigan}. 
The goal  of an operational theory is merely to specify the probabilities $p(k|M,P,T)$ 
of different outcomes $k\in\mathcal{K}_M$ that may result from a measurement procedure 
$M\in\mathcal{M}$ given a particular preparation procedure $P\in\mathcal{P}$, and 
a particular transformation procedure $T\in\mathcal{T}$; where $\mathcal{M}$, $\mathcal{P}$ and 
$\mathcal{T}$ respectively denote the sets of measurement procedures, preparation procedures and 
transformation procedures; $\mathcal{K}_M$ denotes the set of measurement results for 
the measurement M. When there is no transformation procedure, we simply have $p(k|M,P)$. The only 
restrictions on $\{p(k|M,P)\}_{k\in\mathcal{K}_M}$ is that all of them are non negative and 
$\sum_{k\in\mathcal{K}_M}p(k|M,P)=1$ $\forall$ M, P.  As an example, in an operational formulation 
of QT, every preparation P is associated with a density operator $\rho$ on Hilbert space, and every measurement M is associated with a positive operator valued measure (POVM) $\{E_k|~E_k\ge0~\forall~k~\mbox{and}~\sum_kE_k=I\}$. The probability of obtaining outcome $k$ is given by the generalized Born rule, $p(k|M,P)=\mbox{Tr}(E_k\rho)$.

Whereas an operational theory does not tell anything about \emph{physical state} of the system, in an ontological model of an operational theory, the primitives of description are the properties of microscopic systems. A preparation procedure is assumed to prepare a system with certain properties and a measurement procedure is assumed to reveal something about those properties. A complete specification of the properties of a system is referred to as the ontic state of that system. In an ontological model for QT a particular preparation method (context) $\mathcal{C}_{\rho}$ of the quantum state $\rho$ actually yields a probability distribution $\mu(\lambda|\rho,\mathcal{C}_{\rho})$ over the ontic state $\lambda\in\Lambda$, where $\Lambda$ denotes the ontic state space. $\mu(\lambda|\rho,\mathcal{C}_{\rho})$ is called the \emph{epistemic state} associated with $\rho$ and it must satisfy:
\begin{equation}
\int_{\Lambda}\mu(\lambda|\rho,\mathcal{C}_{\rho})d\lambda=1~~\forall~\rho~\mbox{and}~\mathcal{C}_{\rho}.
\end{equation} 
The probability of obtaining the $k$'th outcome is given by the response function, 
$\xi(k|\lambda,\mathcal{C}_{E_k})$, where $\mathcal{C}_{E_k}$ denotes the particular 
measurement method (context) used. When contextuality \footnote{The notion of contextuality is
discussed more elaborately in the next section.} is not relevant, 
the notations $\mathcal{C}_{\rho}$ and $\mathcal{C}_{E_k}$ will be omitted. 

For the ontological model to reproduce quantum predictions for 
a quantum mechanical state $\rho$ , the following must be satisfied
\begin{equation}\label{rho}
\int_{\Lambda} \xi(k|\lambda)\mu(\lambda|\rho) d\lambda = \mbox{Tr}(\rho E_k)
\end{equation} 
In special cases, when pure quantum states are associated with 
preparation and measurements are projective, the above relation reduces to
\begin{equation}\label{psi}
\int_{\Lambda} \xi(\phi|\lambda)\mu(\lambda|\psi) d\lambda = |\langle\phi|\psi\rangle|^2.
\end{equation}

An ontological model is considered to be \emph{$\psi$-ontic} if the specification of 
the ontic state $\lambda$ uniquely determines the quantum state $\psi$ \cite{harrigan}. For 
this to be true, it is necessary that the preparations of any pair of different quantum states, 
$\psi$ and $\phi$, should yield ontic state distributions whose supports, 
$\Lambda_{\psi}$ and $\Lambda_{\phi}$, do not overlap. Hence, 
the epistemic states associated with distinct quantum states are 
completely non-overlapping in a $\psi$-ontic model. In other words, 
different quantum states pick out disjoint regions of $\Lambda$.
A variation of $\psi$, therefore, implies a variation of reality.

If an ontological model fails to be \emph{$\psi$-ontic}, then it is said to 
be \emph{$\psi$-epistemic}, i.e., the class $\psi$-epistemic was merely defined 
as the complement of $\psi$-ontic. In a $\psi$-epistemic model, distinct 
quantum states are consistent with the same state of reality, i.e., variation of $\psi$ does 
not necessarily imply a variation of reality. It is in
this sense, quantum states are judged epistemic in such models.

There is possibility of \emph{trivial $\psi$-epistemic} models, 
such as ones for which there is only a single pair of states, 
$|\psi\rangle$ and $|\phi\rangle$, for which $\Lambda_{\phi}$ 
and $\Lambda_{\psi}$ overlap. However, there are 
\emph{nontrivial $\psi$-epistemic} models. Indeed, 
Aaronson et al. \cite{aaronson} have shown that there are 
maximally-nontrivial $\psi$-epistemic models, for which the 
supports $\Lambda_{\phi}$ and $\Lambda_{\psi}$ overlap for 
every non-orthogonal pair, $|\psi\rangle$ and $|\phi\rangle$. 
These models can be constructed for any finite dimensional Hilbert space. 
Very recently Maroney has introduced the concept of \emph{degree of epistemicity} 
of an ontological model \cite{maroney1}. In the ontological model the following 
elementary relation holds:
\begin{eqnarray}
\int_{\Lambda_{\phi}}\mu(\lambda|\psi)d\lambda
&=&\int_{\Lambda_{\phi}}\xi(\phi|\lambda)\mu(\lambda|\psi)d\lambda\nonumber\\
&\le&\int_{\Lambda}\xi(\phi|\lambda)\mu(\lambda|\psi)d\lambda=|\langle\phi|\psi\rangle|^2
\end{eqnarray}  
The first line follows from the fact that an ontic state $\lambda$ that is compatible with the state preparation $|\phi\rangle$ must assign value $1$ to the response function $\xi(\phi|\lambda)$. The above equation can be expressed as:
\begin{equation}
\int_{\Lambda_{\phi}}\mu(\lambda|\psi)d\lambda=f(\phi,\psi)|\langle\phi|\psi\rangle|^2,
\end{equation} 
where $0\le f(\phi,\psi)\le 1$. In a \emph{$\psi$-ontic} theory, $f(\phi,\psi)= 0$ for every pair of
different quantum states. An ontological model will be called \emph{maximally $\psi$-epistemic} 
if $f(\phi,\psi)= 1$ for all $|\phi\rangle$ and $|\psi\rangle$, otherwise the model will be 
called \emph{nonmaximally $\psi$-epistemic}. The Kochen-Specker model for 
a spin-$\frac{1}{2}$ system is a nice example of a maximally $\psi$-epistemic model. However, such a
model is not possible for Hilbert spaces of dimensions three or more \cite{maroney1}.

\section{Preparation contextuality for mixed qubit states}\label{sec3}

The traditional notion of \emph{contextuality} \cite{kochen} was generalized by Spekkens 
in \cite{spek05} and has been discussed more elaborately in Ref.\cite{rudolph}. While the 
well known notion of KS-contextuality\cite{kochen} is applicable for sharp (projective/ Von-neumann) 
measurements of QT only, this generalized notion of contextuality is applicable to 
an arbitrary operational theory rather than just to QT and to arbitrary experimental 
procedures (preparation procedure, transformation procedure as well as unsharp measurement procedure) 
rather than just to sharp measurements and also it is applicable to a broad class of ontological 
models of QT rather than just to deterministic hidden variable models. 

As mentioned in the previous section, the role of an operational theory is merely 
to specify the probabilities $p(k|\mathcal{M},\mathcal{P})$ of different outcomes $k$ that 
may result from a measurement procedure $\mathcal{M}$ given a particular preparation procedure 
$\mathcal{P}$. Given the rule for 
determining probabilities of outcomes, one can define a notion of equivalence among experimental 
procedures. Two preparation procedures are deemed equivalent if they yield the same long-run statistics for every possible measurement procedure, that is, $\mathcal{P}$ is equivalent to $\mathcal{P'}$ if:
\begin{equation}
p(k|\mathcal{M},\mathcal{P})=p(k|\mathcal{M},\mathcal{P'})~\forall~\mathcal{M}.
\end{equation}
It might happen that the mere specification of the equivalence class of a procedure does not specify
the procedure completely. The set of features of an experimental procedure which are not specified
by specifying the equivalence class is called the \emph{context} of the experimental procedure. An
ontological model is called \emph{noncontextual} if the representation of every experimental procedure
in the model depends only on the equivalence class and not on its contexts.

In an ontological model for QT, an equivalence class of preparation procedures is associated with 
a density operator $\rho$. Hence, the model will be called \emph{preparation non-contextual} if 
it associates a single epistemic state $\mu(\lambda|\rho)$ with a given density operator, $\rho$, 
regardless of its context of preparation. Conversely a model is said to be 
\emph{preparation contextual} if the epistemic state that it assigns to $\rho$ depends 
on the context of its preparation, i.e. there exist different contexts of preparation
$\mathcal{C}_{\rho}$ and $\mathcal{C}'_{\rho}$ giving rise to 
the same density operator $\rho$ such that 
$\mu(\lambda|\rho,\mathcal{C}_{\rho})\ne\mu(\lambda|\rho,\mathcal{C}'_{\rho})$.

Spekkens proved the preparation contextuality for completely mixed state of a qubit \cite{spek05}. 
In the following, we extend this result for any mixed qubit state.

\textbf{Lemma 1}: \emph{Any mixed state of a qubit is preparation contextual}. 

To prove the lemma, we adapt the proof of preparation contextuality 
for completely mixed state of a qubit as in Ref. \cite{spek05}. We write below 
a couple of features of the representations of preparation procedures 
in an ontological model as we will use them in proving the lemma.

\begin{figure}[h!]
\includegraphics[width=8cm, height=11cm]{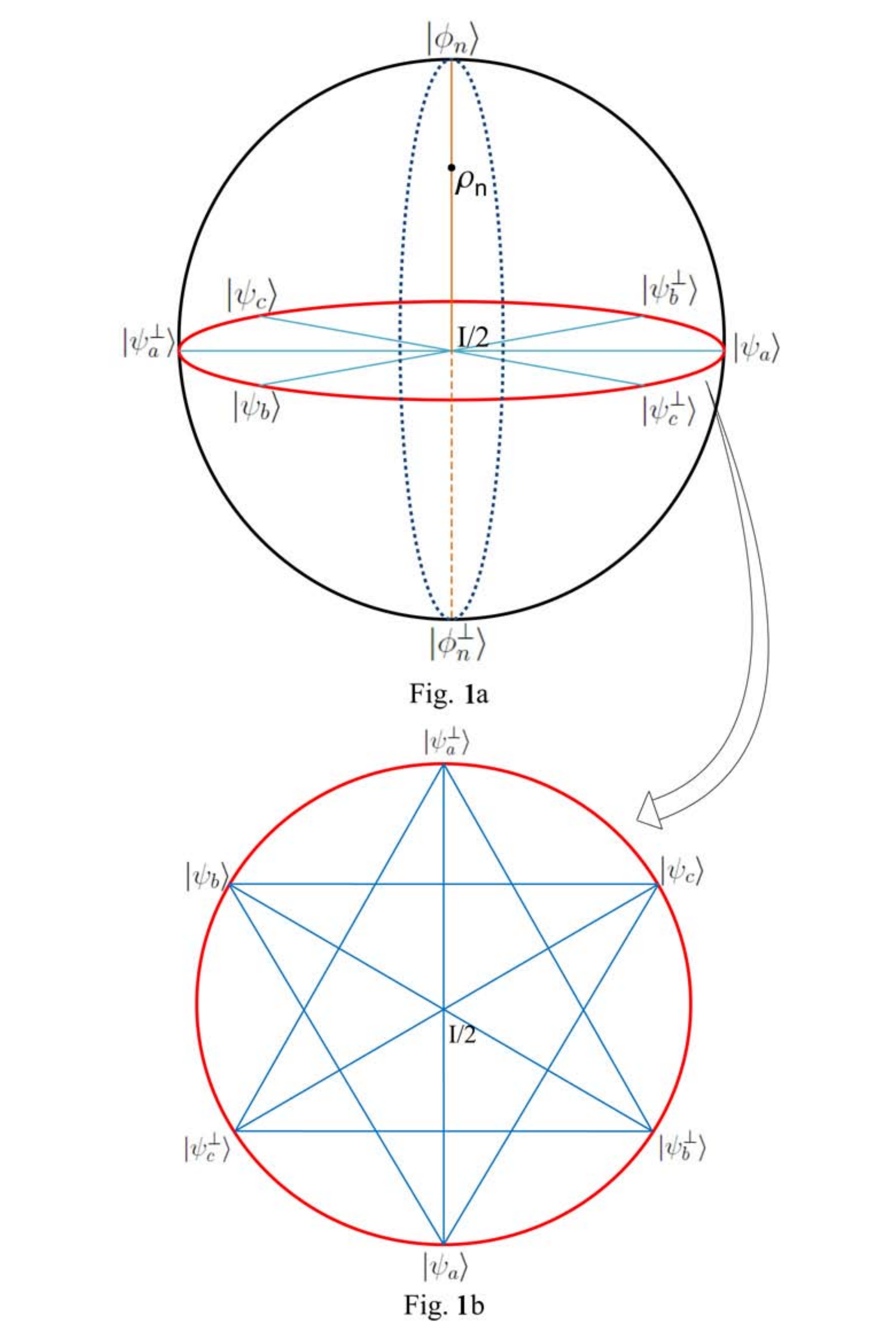}
\caption{ The six decompositions of the mixed state $\rho_n$ of a qubit as given in Eqs.(\ref{first})-(\ref{sixth}).}
\label{fig2}
\end{figure}

\begin{enumerate}
\item[$\bullet$] \emph{Feature 1:} If two preparation procedures are distinguishable with certainty in a single-shot measurement, then their associated probability distributions are non-overlapping, i.e.,
\begin{equation}
\Lambda_{\psi}\cap\Lambda_{\phi}=0~\mbox{for~all~orthogonal~pair}~|\phi\rangle~\mbox{and}~|\psi\rangle
\end{equation}

\item[$\bullet$] \emph{Feature 2:} A convex combination of preparation procedures is represented within an ontological model by a convex sum of the associated probability distributions.
\end{enumerate}
\textbf{Proof:} A mixed qubit $\rho_n =\frac{1}{2}(I+\vec{n}.\vec{\sigma})$, with $0\le |\vec{n}|(=q)<1$, can have several decompositions. A different preparation procedure is associated with every such decomposition. Consider the following decompositions of a mixed state $\rho_n$ of a qubit:
\begin{eqnarray} 
\rho_n &=&\frac{1-q}{2}|\phi_n^{\perp}\rangle\langle\phi_n^{\perp}|+\frac{1+q}{2}|\phi_n\rangle\langle\phi_n|\label {first}\\
&=& \frac{1-q}{2}(|\psi_a\rangle\langle\psi_a|+|\psi_a^{\perp}\rangle\langle\psi_a^{\perp}|)+q|\phi_n\rangle\langle\phi_n|\label {second}\\
&=& \frac{1-q}{2}(|\psi_b\rangle\langle\psi_b|+|\psi_b^{\perp}\rangle\langle\psi_b^{\perp}|)+q|\phi_n\rangle\langle\phi_n|\label {third}\\
&=& \frac{1-q}{2}(|\psi_c\rangle\langle\psi_c|+|\psi_c^{\perp}\rangle\langle\psi_c^{\perp}|)+q|\phi_n\rangle\langle\phi_n|\label {fourth}\\
&=& \frac{1-q}{3}(|\psi_a\rangle\langle\psi_a|+|\psi_b\rangle\langle\psi_b|+|\psi_c\rangle\langle\psi_c|)\nonumber\\
&& +q|\phi_n\rangle\langle\phi_n|\label {fifth}\\
&=& \frac{1-q}{3}(|\psi_a^{\perp}\rangle\langle\psi_a^{\perp}|+|\psi_b^{\perp}\rangle\langle\psi_b^{\perp}|+|\psi_c^{\perp}\rangle\langle\psi_c^{\perp}|)\nonumber\\
&& +q|\phi_n\rangle\langle\phi_n|\label {sixth}.
\end{eqnarray}
where $|\phi_n\rangle\langle\phi_n|=\frac{1}{2}(I+\hat{n}.\vec{\sigma})$ and the vectors $|\psi_a\rangle,|\psi_b\rangle,|\psi_c\rangle$ are chosen from equatorial plane perpendicular to $\hat{n}$ in such a manner that the line joining the points corresponding to $|\psi_a\rangle,|\psi_a^{\perp}\rangle$ makes an angle $60^0$ with the other two lines containing $|\psi_b\rangle,|\psi_b^{\perp}\rangle$ and $|\psi_c\rangle,|\psi_c^{\perp}\rangle$ respectively (Fig.\ref{fig2}).  

As whenever two density operators are orthogonal in the vector space of operators, the associated preparation procedures can be distinguished with certainty in a single shot measurement and as whenever two preparation procedures are distinguishable with certainty in a single shot measurement, their associated probability distribution are non overlapping  (\emph{Feature 1}), we have
\begin{eqnarray}
\mu(\lambda|\phi_n)\mu(\lambda|\phi^{\perp}_n) &=& 0 \label{ortho1}\\
\mu(\lambda|\psi_a)\mu(\lambda|\psi_a^{\perp}) &=& 0 \label{ortho2}\\
\mu(\lambda|\psi_b)\mu(\lambda|\psi_b^{\perp}) &=& 0\label{ortho3} \\
\mu(\lambda|\psi_c)\mu(\lambda|\psi_c^{\perp}) &=& 0\label{ortho4}
\end{eqnarray}
The above six decompositions (Eqs.(\ref{first})-(\ref{sixth})) of $\rho_n$ are associated with six different preparation procedures $\mathcal{C}_{\phi_n^{\perp}\phi_n}$, $\mathcal{C}_{\psi_a\psi_a^{\perp}\phi_n}$,....,$\mathcal{C}_{\psi_a^{\perp}\psi_b^{\perp}\psi_c^{\perp}\phi_n}$ respectively.

As in an ontological model a convex combination of preparation procedures is represented by a convex sum of associated probability distributions (\emph{Feature 2}), so 
\begin{eqnarray}\label{mu}
\mu(\lambda|\rho_n,\mathcal{C}_{\phi_n^{\perp}\phi_n}) &=& \frac{1-q}{2}\mu(\lambda|\phi_n^{\perp})+\frac{1+q}{2}\mu(\lambda|\phi_n) \label{mu1}\\
\mu(\lambda|\rho_n,\mathcal{C}_{\psi_a\psi_a^{\perp}\phi_n}) &=&  \frac{1-q}{2}[\mu(\lambda|\psi_a)+\mu(\lambda|\psi_a^{\perp})]+q\mu(\lambda|\phi_n)\label{mu2} \\
\mu(\lambda|\rho_n,\mathcal{C}_{\psi_b\psi_b^{\perp}\phi_n}) &=&  \frac{1-q}{2}[\mu(\lambda|\psi_b)+\mu(\lambda|\psi_b^{\perp})]+q\mu(\lambda|\phi_n)\label{mu3} \\
\mu(\lambda|\rho_n,\mathcal{C}_{\psi_c\psi_c^{\perp}\phi_n}) &=&  \frac{1-q}{2}[\mu(\lambda|\psi_c)+\mu(\lambda|\psi_c^{\perp})]+q\mu(\lambda|\phi_n)\label{mu4} \\
\mu(\lambda|\rho_n,\mathcal{C}_{\psi_a\psi_b\psi_c\phi_n}) &=&  \frac{1-q}{3}[\mu(\lambda|\psi_a)+\mu(\lambda|\psi_b)+\mu(\lambda|\psi_c)]\nonumber\\ &&+q\mu(\lambda|\phi_n)\label{mu5}\\
\mu(\lambda|\rho_n,\mathcal{C}_{\psi_a^{\perp}\psi_b^{\perp}\psi_c^{\perp}\phi_n}) &=&  \frac{1-q}{3}[\mu(\lambda|\psi_a^{\perp})+\mu(\lambda|\psi_b^{\perp})+\mu(\lambda|\psi_c^{\perp})]\nonumber\\
&&+q\mu(\lambda|\phi_n)\label{mu6}
\end{eqnarray}
Let us denote the the support of $\mu(\lambda|\rho_n)$ by $\Lambda_{\rho_n}$, i.e., $\Lambda_{\rho_n}=\{\lambda\in\Lambda~|~\mu(\lambda|\rho_n)>0\}$. As mentioned before, the assumption of preparation non-contextuality demands the distribution of ontic state $\lambda$ associated with a preparation procedure to depend only upon the density matrix $\rho_n$, thus the following equations should also get satisfied
\begin{eqnarray} 
\mu(\lambda|\rho_n) &=& \frac{1-q}{2}\mu(\lambda|\phi_n^{\perp})+\frac{1+q}{2}\mu(\lambda|\phi_n) \label{nu1}\\
&=&  \frac{1-q}{2}[\mu(\lambda|\psi_a)+\mu(\lambda|\psi_a^{\perp})]+q\mu(\lambda|\phi_n) \label{nu2} \\
&=&  \frac{1-q}{2}[\mu(\lambda|\psi_b)+\mu(\lambda|\psi_b^{\perp})]+q\mu(\lambda|\phi_n)  \label{nu3}\\
&=&  \frac{1-q}{2}[\mu(\lambda|\psi_c)+\mu(\lambda|\psi_c^{\perp})]+q\mu(\lambda|\phi_n) \label{nu4} \\
&=&  \frac{1-q}{3}[\mu(\lambda|\psi_a)+\mu(\lambda|\psi_b)+\mu(\lambda|\psi_c)] +q\mu(\lambda|\phi_n) \label{nu5}\\
&=&  \frac{1-q}{3}[\mu(\lambda|\psi_a^{\perp})+\mu(\lambda|\psi_b^{\perp})+\mu(\lambda|\psi_c^{\perp})]+q\mu(\lambda|\phi_n) \label{nu6}
\end{eqnarray}

But there is no distribution which is compatible with Eqs.(\ref{ortho1})-(\ref{ortho4}) and Eqs.(\ref{nu1})-(\ref{nu6}). To see this let us denote the probability of occurrence of an $\lambda\in\Lambda_{\rho_n}$ for different preparations by $\mu(\lambda|\phi_n),....,\mu(\lambda|\psi_c^{\perp})$. To satisfy Eq.(\ref{ortho1}), one out of the pair $\mu(\lambda|\phi_n)$ and $\mu(\lambda|\phi_n^{\perp})$ must be zero. Same is true for the pairs $\{\mu(\lambda|\psi_a), \mu(\lambda|\psi_a^{\perp})\}$, $\{\mu(\lambda|\psi_b), \mu(\lambda|\psi_b^{\perp})\}$ and $\{\mu(\lambda|\psi_c), \mu(\lambda|\psi_c^{\perp})\}$ in order to satisfy Eqs.(\ref{ortho2})-(\ref{ortho4}). Thus, in all, we have sixteen different situations:
\begin{enumerate}
\item[(a)] If $\mu(\lambda|\phi_n)=\mu(\lambda|\psi_a)=\mu(\lambda|\psi_b)=\mu(\lambda|\psi_c)=0$, then from Eq.(\ref{nu5}) we have $\mu(\lambda|\rho_n)=0$, a contradiction. 

\item[(b)]If $\mu(\lambda|\phi_n)=\mu(\lambda|\psi_a^{\perp})=\mu(\lambda|\psi_b)=\mu(\lambda|\psi_c)=0$, then from Eq.(\ref{nu2}) and Eq.(\ref{nu5}) it follows that $\frac{1-q}{2}\mu(\lambda|\psi_a)=\frac{1-q}{3}\mu(\lambda|\psi_a)$, which gives $\mu(\lambda|\psi_a)=0$, as $q>0$. Thus we have $\mu(\lambda|\rho_n)=0$, giving a contradiction. Same is true for the cases $\mu(\lambda|\phi_n)=\mu(\lambda|\psi_a)=\mu(\lambda|\psi_b^{\perp})=\mu(\lambda|\psi_c)=0$ and $\mu(\lambda|\phi_n)=\mu(\lambda|\psi_a)=\mu(\lambda|\psi_b)=\mu(\lambda|\psi_c^{\perp})=0$.

\item[(c)]If $\mu(\lambda|\phi_n)=\mu(\lambda|\psi_a^{\perp})=\mu(\lambda|\psi_b^{\perp})=\mu(\lambda|\psi_c)=0$, then from Eq.(\ref{nu4}) and Eq.(\ref{nu6}) it follows that $\frac{1-q}{2}\mu(\lambda|\psi_c^{\perp})=\frac{1-q}{3}\mu(\lambda|\psi_c^{\perp})$ giving a contradiction. Similar hold for the cases $\mu(\lambda|\phi_n)=\mu(\lambda|\psi_a^{\perp})=\mu(\lambda|\psi_b)=\mu(\lambda|\psi_c^{\perp})=0$ and $\mu(\lambda|\phi_n)=\mu(\lambda|\psi_a)=\mu(\lambda|\psi_b^{\perp})=\mu(\lambda|\psi_c^{\perp})=0$.

\item[(d)]If $\mu(\lambda|\phi_n)=\mu(\lambda|\psi_a^{\perp})=\mu(\lambda|\psi_b^{\perp})=\mu(\lambda|\psi_c^{\perp})=0$, then from Eq.(\ref{nu6}) gives a contradiction. 

\item[(e)] If $\mu(\lambda|\phi_n^{\perp})=\mu(\lambda|\psi_a)=\mu(\lambda|\psi_b)=\mu(\lambda|\psi_c)=0$ then Eq.(\ref{nu1}) and Eq.(\ref{nu5}) imply $\frac{1+q}{2}\mu(\lambda|\phi_n)=q\mu(\lambda|\phi_n)$ giving a contradiction.

\item[(f)] If $\mu(\lambda|\phi_n^{\perp})=\mu(\lambda|\psi_a^{\perp})=\mu(\lambda|\psi_b)=\mu(\lambda|\psi_c)=0$ then Eq.(\ref{nu2}) and Eq.(\ref{nu5}) imply $\frac{1-q}{2}\mu(\lambda|\psi_a)+q\mu(\lambda|\phi_n)=\frac{1-q}{3}\mu(\lambda|\psi_a)+q\mu(\lambda|\phi_n)$  which further implies $\mu(\lambda|\psi_a)=0$. Thus the case reduces to the case (e) and gives contradiction. Similar analysis hold for the cases $\mu(\lambda|\phi_n^{\perp})=\mu(\lambda|\psi_a)=\mu(\lambda|\psi_b^{\perp})=\mu(\lambda|\psi_c)=0$ and $\mu(\lambda|\phi_n^{\perp})=\mu(\lambda|\psi_a)=\mu(\lambda|\psi_b)=\mu(\lambda|\psi_c^{\perp})=0$.

\item[(g)] If $\mu(\lambda|\phi_n^{\perp})=\mu(\lambda|\psi_a^{\perp})=\mu(\lambda|\psi_b^{\perp})=\mu(\lambda|\psi_c)=0$ then Eq.(\ref{nu4}) and Eq.(\ref{nu6}) give $\frac{1-q}{2}\mu(\lambda|\psi_c^{\perp})+q\mu(\lambda|\phi_n)=\frac{1-q}{3}\mu(\lambda|\psi_c^{\perp})+q\mu(\lambda|\phi_n)$  which implies $\mu(\lambda|\psi_c^{\perp})=0$. This case reduces to the following case (case (h)) and gives contradiction. Similar analysis hold for the cases $\mu(\lambda|\phi_n^{\perp})=\mu(\lambda|\psi_a^{\perp})=\mu(\lambda|\psi_b)=\mu(\lambda|\psi_c^{\perp})=0$ and $\mu(\lambda|\phi_n^{\perp})=\mu(\lambda|\psi_a)=\mu(\lambda|\psi_b^{\perp})=\mu(\lambda|\psi_c^{\perp})=0$.

\item[(h)] If $\mu(\lambda|\phi_n^{\perp})=\mu(\lambda|\psi_a^{\perp})=\mu(\lambda|\psi_b^{\perp})=\mu(\lambda|\psi_c^{\perp})=0$ then Eq.(\ref{nu1}) and Eq.(\ref{nu6}) give $\frac{1+q}{2}\mu(\lambda|\phi_n)=q\mu(\lambda|\phi_n)$ which implies $\mu(\lambda|\phi_n)=0$ (as $q<1$) and the case boils down to the case (d).

\end{enumerate} 
The above argument holds for any $\lambda\in\Lambda_{\rho_n}$. Thus we conclude that for the density matrix $\rho_n$, preparation noncontextual assignment of ontic state $\lambda$ is not possible.

Unlike the various proofs of measurement contextuality using different resolutions of identity 
in terms of one dimensional projectors \cite{peres,mermin,cabello}, the meaning of preparation 
contextuality needs further elaboration. Two extreme cases that may be implied by preparation 
contextuality are as follows: 
\begin{itemize}
\item[(1)] Distribution of the ontic variables $\lambda$ for a mixed state are different for 
its different preparation procedures while distribution of the ontic variables $\lambda$ is same 
for every preparation of a pure state;
\item[(2)] Distribution of $\lambda$ for pure states may depend on the context created by 
its different preparation procedures but the distribution of the ontic variables $\lambda$ for 
the mixed state remains same for all preparation procedures.
\end{itemize}

To explore the role of preparation contextuality in demonstrating nonlocality of entangled states, we
consider the above two cases separately. In the first case, the nonlocality of
any pure entangled state of two-qubits in some $\psi$-epsitemic models could follow
from Schrodinger-Gisin-Hughston-Jozsa-Wootters (GHJW) steering \cite{Schro,hjw,ghjw2}. 
To see this explicitly, consider a situation where two far separated parties Alice and Bob share a non maximally pure entangled state of two qubits. The reduced state of Bob's system is then 
a mixed qubit $\rho_n =\frac{1}{2}(I+\vec{n}.\vec{\sigma})$ which can have the above mentioned 
six different decompositions (Eqs.(\ref{first})-(\ref{sixth})). Now, according to the GHJW theorem 
\footnote {This theorem states that every representation of a density matrix
$\rho_B$ can be prepared by acting on a different non-interacting
system A if A and B share a  pure entangled state $|\psi_{AB}\rangle $ such that $\rho_B=Tr_{A} (|\psi_{AB}\rangle 
\langle \psi_{AB}|)$.}, Alice can remotely prepare 
the Bob's system in any of these decompositions (Eqs.(\ref{first})-(\ref{sixth})) and
according to the first implication, these ensembles do not correspond to the 
same probability distribution over ontic states. Thus, the distribution on Bob's side depends 
on Alice's choice of measurement, which implies nonlocality.
In the second case, the conditional distributions of the ontic variable changes for a same pure state when
prepared in different context. Though this may imply some sort of
steering within the ontological model itself, this cannot be used to reveal the nonlocality 
as the unconditional distribution of the ontic variable remains the same.
Hence the preparation contextuality of $\psi$-epsitemic models as shown above, cannot be used to demonstrate nonlocality of entangled states. Interestingly, in the next section, we show that there is a way to bypass the second implication to link preparation contextuality with nonlocality of pure entangled states.

\section{Nonlocality of bipartite pure entangled states}\label{sec4}

Liefer and Maroney have demonstrated the nonlocality of a non maximally 
$\psi$-epistemic model without using Bell's inequality \cite{maroney}. 
In order to show this nonlocality, they have first shown 
preparation contextuality of a completely mixed qubit. The nonlocality of a non 
maximally $\psi$-epistemic model has then been demonstrated by the use of the fact
that Alice can remotely prepare different decompositions of Bob's reduced density matrix
when they share a maximally entangled state of two-qubit. 
In the following, we show that this approach of demonstrating nonlocality
by preparation contextuality and steerability, can be extended to demonstrate nonlocality 
of any bipartite pure entangled state provided we consider  non-maximallly $\psi$-epistemic models where for 
every $\psi$, there exists at least one \footnote{The scenario considered by Leifer 
and Maroney is more relaxed as it requires to satisfy 
Eq.(\ref{stronger}) only for a single pair of $\psi$ and $\phi$.} $\phi$ such that 
\begin{equation}\label{stronger}
\int_{\Lambda_{\phi}} \mu(\lambda|\psi) d \lambda <|\langle\phi|\psi\rangle|^2.
\end{equation}   

For convenience, we proceed in two steps. We first show nonlocality of any pure entangled state 
of two-qubit and then generalize the proof to include any bipartite pure entangled state.

\begin{figure}[t!]
\includegraphics[width=6.3cm, height=5cm]{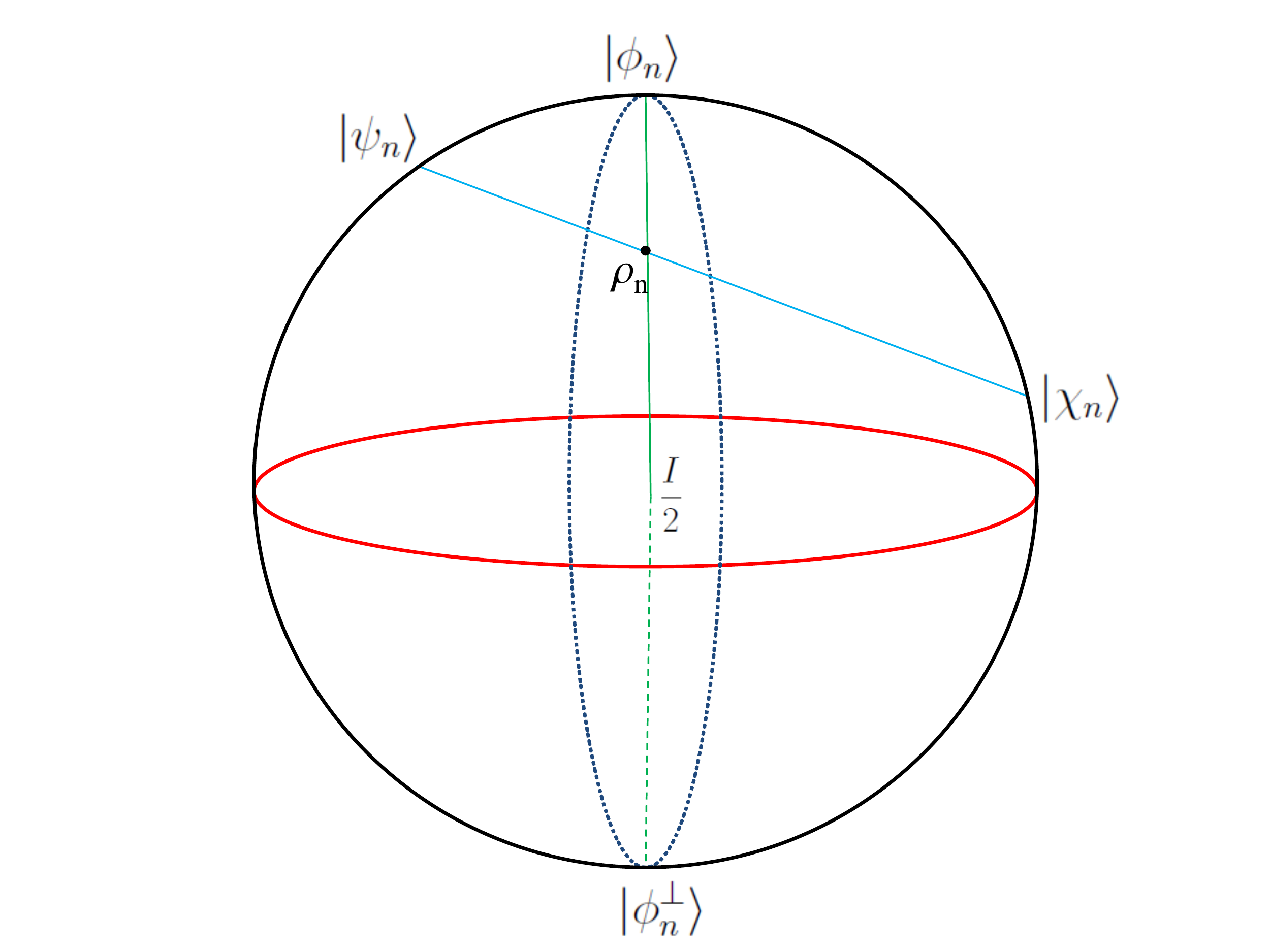}
\caption{The two decompositions of mixed state of a qubit $\rho_n$ corresponding to preparation procedures (I) and (II), where (I): $\rho_n=\frac{1}{2}(\mathbf{1}+\vec{n}.\vec{\sigma})=\frac{1+q}{2}|\phi_n\rangle\langle\phi_n|+\frac{1-q}{2}|\phi_n^{\perp}\rangle\langle\phi_n^{\perp}|$, and (II): $\rho_n=r|\psi_n\rangle\langle\psi_n|+(1-r)|\chi_n\rangle\langle\chi_n|$.}
\label{pic}
\end{figure}

{\bf(a) Nonlocality of pure entangled states of two qubits}: 

Consider the following two preparation procedures (see Fig.\ref{pic}) of a mixed qubit $\rho_n=\frac{1}{2}(I+\vec{n}.\vec{\sigma})$ (with $0\le|\vec{n}|(=q)<1$):
\begin{enumerate}
\item[(I)] Where either of the two preparations $\mathcal{C}_{\phi_{n}}$ and $\mathcal{C}_{\phi_{n}^{\perp}}$ corresponding to orthogonal
quantum mechanical states $|\phi_n\rangle$ and $|\phi_n^{\perp}\rangle$ are implemented with respective probabilities $\frac{1+q}{2}$ and 
$\frac{1-q}{2}$.
\item[(II)] Where the two preparations $\mathcal{C}_{\psi_{n}}$ and $\mathcal{C}_{\chi_{n}}$ corresponding to nonorthogonal
quantum mechanical states $|\psi_n\rangle$ and $|\chi_n\rangle$ are implemented with respective probabilities $r$ and 
$(1-r)$. Here $|\psi_n\rangle$ is chosen in such a manner that relation (\ref{stronger}) holds for the pair $|\psi_n\rangle$ and $|\phi_n\rangle$.
\end{enumerate}
Using (\ref{psi}) for $|\langle\phi_n|\psi_n\rangle|^2$, we get from (\ref{stronger})
\begin{equation}\label{noneed}
\int_{\Lambda_{\phi_n}} \mu(\lambda|\psi_n) d \lambda <\int_{\Lambda} \xi(\phi_n|\mathcal{M},\lambda)\mu(\lambda|\psi_n) d\lambda
\end{equation}
As $ \xi(\phi_n|\mathcal{M},\lambda)=1$, almost everywhere on $\Lambda_{\phi_n}$; hence  
\begin{equation}\label{need}
\int_{\Lambda_{\phi_n}} \xi(\phi_n|\mathcal{M},\lambda)\mu(\lambda|\psi_n) d\lambda <\int_{\Lambda} \xi(\phi_n|\mathcal{M},\lambda)\mu(\lambda|\psi_n) d\lambda
\end{equation}
 This means that there is a set $\Omega$ of ontic states of non zero measure such that:
\begin{itemize}
\item [(i)] $\Lambda_{\phi_n}\cap\Omega=\emptyset$, 
\item[(ii)] $\Omega$ is assigned nonzero probability by $\mathcal{C}_{\psi_n}$,
\item [(iii)] and $\xi(\phi_n|\mathcal{M},\lambda)> 0$ for $\lambda\in \Omega$.
\end{itemize}
In an ontological model, a convex combination of preparation procedures is
represented by a convex sum of the associated probability distributions (\emph {Feature 2} ). 
Hence, if the preparation procedures I and II are represented by $\mathcal{C}_{\phi_{n}\phi_{n}^{\perp}}$
and $\mathcal{C}_{\psi_{n}\chi_{n}}$ respectively, we have
\begin{eqnarray}
\mu(\lambda|\rho,\mathcal{C}_{\phi_{n}\phi_{n}^{\perp}}) &=& \frac{(1+q)}{2} \mu(\lambda|\phi_{n})+\frac{(1-q)}{2} \mu(\lambda|\phi_{n}^{\perp}) \\
\mu(\lambda|\rho,\mathcal{C}_{\psi_{n}\chi_{n}}) &=&  r\mu(\lambda|\psi_{n})+(1-r)\mu(\lambda|\chi_{n})
\end{eqnarray}
The region $\Omega$ is assigned zero probability by $\mathcal{C}_{\phi_{n}}$ as $\Lambda_{\phi_n}\cap\Omega=\emptyset$ and $\mathcal{C}_{\phi_{n}^{\perp}}$ must assign zero probability to any set of ontic states that assigns nonzero probability to $\phi_n$ in a measurement of any orthonormal basis that contains it (i.e. $\xi(\phi_n|\mathcal{M},\lambda)> 0$). Thus, $\Lambda_{\phi_{n}\phi_{n}^{\perp}}\cap\Omega=\emptyset$, where $\Lambda_{\phi_{n}\phi_{n}^{\perp}}=\Lambda_{\phi_{n}}\cup\Lambda_{\phi_{n}^{\perp}}$ is the support of $\mu(\lambda|\rho,\mathcal{C}_{\phi_{n}\phi_{n}^{\perp}})$. Hence the region $\Lambda_{\psi_{n}\chi_{n}}\cap\Omega$ $\left( \mbox{where}~ \Lambda_{\psi_{n}\chi_{n}}=\Lambda_{\psi_{n}}\cup\Lambda_{\chi_n}\right.$ is the support of $\mu(\lambda|\rho,\mathcal{C}_{\psi_{n}\chi_{n}}))$ is assigned zero probability by preparation (I) whereas preparation (II) assigns non zero probability to it. Thus $\mu(\lambda|\rho,\mathcal{C}_{\phi_{n}\phi_{n}^{\perp}})$ and $\mu(\lambda|\rho,\mathcal{C}_{\psi_{n}\chi_{n}})$ are distinct distributions of ontic states of the system. On the contrary, the assumption of preparation non-contextuality demands a distribution associated with a preparation procedure to depend only on the density operator associated with that procedure and not on the particular decomposition of the density matrix \cite{spek05,rudolph}.

This preparation contextuality can be used to show nonlocality of a nonmaximally
pure entangled state of two qubit. Consider a situation where two
far separated parties Alice and Bob share a nonmaximally pure entangled state
of two qubits such that the reduced density matrix of Bob's system is given by
\begin{eqnarray}\label{pre1}
\rho_n &=& \frac{(1+q)}{2} |\phi_n\rangle\langle\phi_n|+\frac{(1-q)}{2} |\phi_n^{\perp}\rangle\langle\phi_n^{\perp}| \\\label{pre2}
&=&  r|\psi_n\rangle\langle\psi_n|+(1-r)|\chi_n\rangle\langle\chi_n|
\end{eqnarray}
According to the Gisin-Hughston-Jozsa-Wootters (GHJW) theorem \cite{ghjw2}, Alice can remotely prepare Bob's system either in (\ref{pre1}) or in (\ref{pre2}). However,
as shown above, these two ensembles cannot correspond to the same
probability distribution over ontic states. Thus, the distribution on Bob's side
depends on Alice's choice of measurement, which implies nonlocality.

{\bf(b)Nonlocality of any bipartite pure entangled state}: 

Here we concentrate on a more general scenario, i.e., we consider any bipartite pure entangled state. It would be sufficient to analyze the following two cases: (1) the bipartite pure entangled state is of Schmidt rank two and (2) the bipartite pure entangled state is of Schmidt rank greater than two.

{\bf Case (1)}: Whenever the entangled state shared between Alice and Bob is of Schmidt rank two, the support of the marginal density matrix is a $2$ dimensional subspace and thus the nonlocality argument reduces to the two qubit scenario discussed just before.

{\bf Case (2)}: Consider a composite system of dimension $d_1\times d_2$. Consider a situation where 
two distantly separated parties Alice and Bob share a pure entangled state of this composite 
system with Schmidt rank $d$, where $2< d\le \mbox{min}\{d_1,d_2\}$. The support of the Bob's reduced density matrix $\rho_B$ is a $d$ dimensional subspace. Let the spectral decomposition of $\rho_B$ is:
\begin{equation} 
\rho_B=\sum_{j=1}^dp_j|\phi_j\rangle\langle\phi_j|,\label{prep1}
\end{equation}
where $\{p_j\}_j$ is a probability distributions and $\langle\phi_i|\phi_j\rangle=\delta_{ij}$. Here, the support of $\rho_B$ is of dimension greater than two, so according to \cite{maroney1}\footnote{In \cite{maroney1}, Maroney  has shown that for every quantum state (in Hilbert spaces of dimension greater than two) there exists another quantum state such that the ontic overlap is strictly less than the quantum overlap and hence the ontological model cannot be maximally $\psi$-epistemic.}
, for the pure state $|\phi_1\rangle$ there exists another state $|\psi_1\rangle$, belonging to the support of $\rho_B$, such that the relation (\ref{stronger}) holds for this pair. Therefore, in this case we need not to assume the stronger form of non maximally $\psi$-epitemicity, rather the requirement is an established fact. Since $|\psi_1\rangle$ belongs to the range of $\rho_B$, so there exists a pure states decomposition of $\rho_B$ containing $|\psi_1\rangle$ as an element of that decomposition \cite{Lahti,Cassinelli}, i.e.,
\begin{equation} 
\rho_B=q_1|\psi_1\rangle\langle\psi_1|+\sum_{k>1}q_k|\psi_k\rangle\langle\psi_k|,\label{prep2}
\end{equation}
where $\{q_k\}_{k=1,2,..}$ is a probability distribution. The above two decompositions of $\rho_B$
(Eq.(\ref{prep1}) and Eq.(\ref{prep2})) correspond to the following two different preparation procedures 
for it respectively: 
\begin{enumerate}
\item[($\alpha$)] The preparations $\{\mathcal{C}_{\phi_j}\}_j$ corresponding to the orthogonal quantum mechanical states $\{|\phi_j\rangle\}_j$ are implemented with respective probabilities $\{p_j\}_j$.  

\item[($\beta$)] The preparations $\{\mathcal{C}_{\psi_k}\}_k$ corresponding to the quantum mechanical states $\{|\psi_k\rangle\}_k$ are implemented with  probabilities $\{q_k\}_k$ respectively. 
\end{enumerate} 
Denote the support of the ontic states corresponding to the preparation procedures $(\alpha)$ and $(\beta)$ as $\Lambda_{\{{\phi_j}\}_j}$ and $\Lambda_{\{{\psi_k}\}_k}$, respectively. Arguing in a manner similar to the two-qubit scenario, it can be shown that there exists a set $\Omega$ of ontic states of non zero measure such that $\Lambda_{\{{\phi_j}\}_j}\cap\Omega=\emptyset$ whereas $\Lambda_{\{{\psi_k}\}_k}\cap\Omega$ is of non zero measure.
Hence the region $\Lambda_{\{\psi_k\}_k}\cap\Omega$ is assigned zero probability by preparation $(\alpha)$ whereas preparation $(\beta)$ assigns non zero probability to it. Which implies that the mixed state $\rho_B$ is preparation contextual. According to the GHJW theorem, Alice can remotely prepare Bob's system either in (\ref{prep1}) or in (\ref{prep2}). Thus, the distribution on Bob's side depends on Alice's choice of measurement, which implies  nonlocality.

\section{Conclusion}
The concept of steering was first introduced by Schr\"{o}dinger
in 1935 \cite{Schro}. Though this concept was disturbing to him, 
but it has no direct implication for nonlocality. Steering has 
attracted much attention in recent years \cite{wiseman1,wiseman2,wiseman3}. 
Whereas the nonlocality of $\psi$-ontic models in presence of steering 
seems more obvious \cite{harrigan}, the role of steering in revealing 
nonlocality of various $\psi$-epistemic models is not so clear. 
It seems that nonlocality of a general $\psi$-epistemic 
model will not follow from steering due to the possibility of 
the second implication of preparation contextuality 
(mentioned in Section-\ref{sec3}). But, interestingly, in a recent
development, it has been shown that just a little loss of 
epistemicity (in the sense that overlap of the ontic supports 
of at least one pair of nonorthogonal pure states is strictly less than the quantum overlap) 
makes nonlocality of such models to follow from steering \cite{maroney}. Extension of 
this loss of epistemicity (i.e., for every pure state there exists at least another nonorthogonal 
pure state such that overlap of their ontic supports is strictly less than the quantum overlap) 
makes it possible to prove even Gisin-like theorem \cite{nicholas} 
for a two-qubit pure entangled state by using steering. We have also established the nonlocality of a bipartite pure entangled state almost in the same spirit as Gisin's theorem but the proof (without any further assumption on epestemic character of the underlying ontological model) works for states with Schmidt rank greater than two. Whether this kind of nonlocality proof can be extended to mixed entangled state or to a class of mixed entangled states remains a gray area of further research.

\section*{Note added}
Recently, M. S. Leifer have written a review on $\psi$-ontology theorems \cite{review}. Among many open questions listed in there one is concerned about preparation contextuality proof for any mixed quantum state. This question is addressed in our paper and we have answered affirmatively.   

\section*{Acknowledgment}

It is our great pleasure to acknowledge G. Kar for various discussions, suggestions and help in proving the results. 
MB acknowledges private communications with M. S. Leifer about stronger non-maximal $\psi$-epistemicity. 
Discussion with S. Ghosh is gratefully acknowledged. SKC is thankful to the Physics and Applied Mathematics Unit, 
Indian Statistical Institute, Kolkata where a major part of this work was done when he was
visiting the unit. AM acknowledges support from the CSIR project
09/093(0148)/2012-EMR-I. SKC acknowledges support from CSIR, Govt. of India. We also like to thank an anonymous referee 
for useful suggestions.

\end{document}